\documentclass{article}
\PassOptionsToPackage{numbers, compress}{natbib}
\usepackage[preprint]{neurips_2023}

\usepackage[utf8]{inputenc} 
\usepackage[T1]{fontenc}    
\usepackage{hyperref}       
\usepackage{url}            
\usepackage{booktabs}       
\usepackage{amsfonts}       
\usepackage{nicefrac}       
\usepackage{microtype}      
\usepackage[dvipsnames]{xcolor}         
\usepackage{enumerate}
\usepackage{enumitem}
\usepackage{amsmath,amsfonts,amssymb}
\usepackage{graphicx}
\usepackage{multirow}
\graphicspath{ {./fig/final/} }

\newcommand{\cn}{{\textit{AttentionLego}}}

\title{\textit{AttentionLego}: An Open-Source Building Block For Spatially-Scalable Large Language Model Accelerator With Processing-In-Memory Technology}

\author{%
Rongqing~Cong\thanks{All authors contributed equally, listed in alphabetical order by last names.} , 
Wenyang~He$^*$, 
Mingxuan~Li$^*$, 
Bangning~Luo$^*$, 
Zebin~Yang$^*$, \\
\textbf{Yuchao~Yang}$^\dagger$,
\textbf{Ru~Huang}$^\dagger$,
\textbf{Bonan~Yan}\thanks{Corresponding authors:Yuchao Yang (yuchaoyang@pku.edu.cn), Ru Huang (ruhuang@pku.edu.cn) and Bonan Yan (bonanyan@pku.edu.cn).}
\\
Peking University
\\\url{https://bonany.cc/attentionlego} \\
}

\begin{document}

\maketitle

\begin{abstract}
Large language models (LLMs) with Transformer architectures have become phenomenal in natural language processing, multimodal generative artificial intelligence, and agent-oriented artificial intelligence.
The self-attention module is the most dominating sub-structure inside Transformer-based LLMs.
Computation using general-purpose graphics processing units (GPUs) inflicts reckless demand for I/O bandwidth for transferring intermediate calculation results between memories and processing units.
To tackle this challenge, this work develops a fully customized vanilla self-attention accelerator, {\cn}, as the basic building block for constructing spatially expandable LLM processors.
{\cn} provides basic implementation with fully-customized digital logic incorporating Processing-In-Memory (PIM) technology. It is based on PIM-based matrix-vector multiplication and look-up table-based Softmax design. The open-source code is available online: \url{https://bonany.cc/attentionleg}.
\end{abstract}

\section{Introduction}

The Transformer architecture has attracted significant attention due to its exceptional performance in a variety of natural language processing, vision, and multimodal generative tasks~\cite{touvron_llama_2023-1,touvron_llama_2023,workshop_bloom_2023,dey_cerebras-gpt_2023,dao_flashattention_2022,black_gpt-neox-20b_2022,li_textbooks_2023,biderman_pythia_2023}. The Transformer was first introduced in 2017 introducing full integration of self-attention mechanism~\cite{vaswani_attention_2023}. It can model complex relationships between different parts of a sequence, making it an ideal choice for handling long-range dependencies and capturing contextual information in sequential signal processing. Experiments have shown that tiling self-attention beyond a specific scale leads to emergent abilities of large language models (LLMs), for example, performing planning, arithmetic, summarizing messages, etc. 

The increasing demand for efficient, intelligent devices and systems highlights the importance of building Large Language Model (LLM) accelerators. LLMs are becoming a key component in Artificial Intelligence and the Internet of Things (AIoT) to enable the integration of natural language processing (NLP) capabilities into various Internet of Things (IoT) applications, allowing for more intuitive and user-friendly interfaces~\cite{yan_resistive_2019}. However, training and deploying LLMs are computationally intensive and cause unreasonable carbon emissions, making it challenging to scale them to meet the demands of IoT devices and systems. Besides, the complex computing mechanism of the self-attention module calls for primers with source code, especially for hardware designers who need a starting point for LLM accelerator design.  

\begin{figure}[h]
	\centering
	\includegraphics[width=0.85\columnwidth]{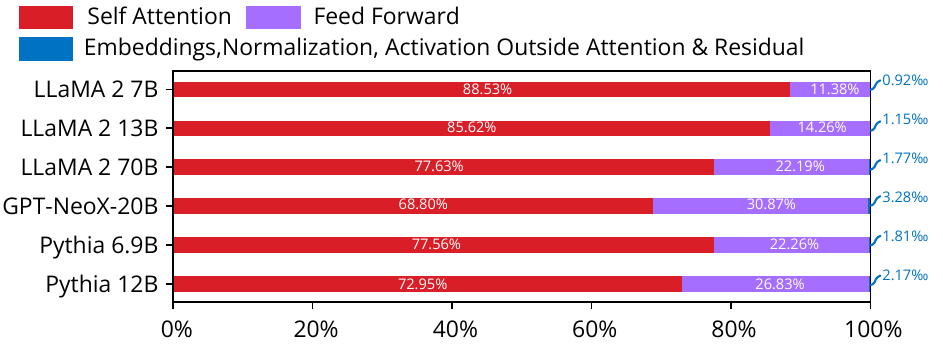}
	\caption{Operation number breakdown for popular large language models. Self-attention module dominates the operation counts in LLMs.1 Multiply-Accumulate (MAC) counts 2 operations. We unify the operation counts for floating-point numbers and integers.}
	\label{fig:llm_ops}
\end{figure}

Realization of LLM accelerators should focus on implementing self-attention modules because the self-attention modules occupy over 68\% of operations in the prevailing LLM architectures (as shown in Fig.~\ref{fig:llm_ops})~\cite{touvron_llama_2023,touvron_llama_2023-1,dey_cerebras-gpt_2023,black_gpt-neox-20b_2022,biderman_pythia_2023,li_textbooks_2023}. With this observation, this work develops a fully-customized vanilla self-attention accelerator, {\cn}. It aims to provide a fundamental building block for constructing spatially expandable LLM processors.
{\cn} implements hardware computation for self-attention with fully customized digital logic incorporating Processing-In-Memory (PIM) technology to boost the computing efficiency. It is based on PIM-based matrix-vector multiplication and look-up-table-based Softmax design. 
This work significantly improves the performance and efficiency of LLMs, making them more accessible to developers and users.

\section{Preliminaries}
\subsection{Processing-In-Memory Technology}

\begin{figure}[h]
	\centering
	\includegraphics[width=1\columnwidth]{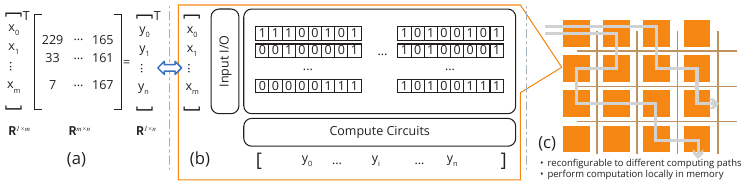}
	\caption{Processing-in-memory macro to perform \textit{in situ} general matrix-vector multiplication.}
	\label{fig:pim}
\end{figure}

Processing-in-memory (PIM) technology is a game-changing innovation that integrates processing units and memory on the same physical chip~\cite{yan_rram-based_2019}.
By collocating memory and processing units, PIM technology eliminates the need for data transfers between the processor and memory, significantly reducing latency and improving performance in various scientific and engineering applications. One application where PIM technology can profoundly impact is matrix-vector multiplication, a fundamental operation in LLMs. According to Fig.~\ref{fig:llm_ops}, the primary operations are self-attention and feed forward layers. Both heavily rely on storage and matrix multiplication, which are the key features offered by PIM. Implementation for the feed forward using PIM has been intensely investigated. Therefore, this work centers on the self-attention module part.

Fig.~\ref{fig:pim} illustrates the principle of using PIM to execute in-memory matrix multiplications. General matrix multiplication can be divided into matrix-vector multiplication (Fig.~\ref{fig:pim}(a)). The deep learning network parameters (synaptic weights) are pre-loaded into the PIM macro array. With the help of PIM peripheral circuits, an input vector is fed into the PIM macro and interacts with the parameters stored in the PIM macro (as depicted in Fig.~\ref{fig:pim}(b)). 
Such PIM macro can be tiled to a spatially expandable architecture (Fig.~\ref{fig:pim}(c)) to store all the parameters/weights used in deep learning networks.
Further, the parallelism of the matrix multiplication per operation is tunable by choosing different design parameters for the PIM macro. For example, to turn on 4, 8, 16-word lines of the PIM macro at each computing step to provide different computing throughput by comprising power consumption and chip area~\cite{yan_resistive_2019,yan_rram-based_2019}. 
In this way, PIM macros, with the help of peripheral digital logic controlling circuits, form accelerators that can exploit the inherent parallelism in matrix-vector multiplication to achieve even higher speeds and lower latencies. This work aims to answer the question of how to utilize the primary matrix-vector multiplication operator(realized by the open-source PIM behavioral model from \url{https://bonany.gitlab.io/pis}~\cite{yan_pislib_2024}) to construct a vanilla (baseline) design for the self-attention module in LLMs. 

\subsection{Self-Attention Module in Large Language Models}

\begin{figure}[h]
	\centering
	\includegraphics[width=0.9\columnwidth]{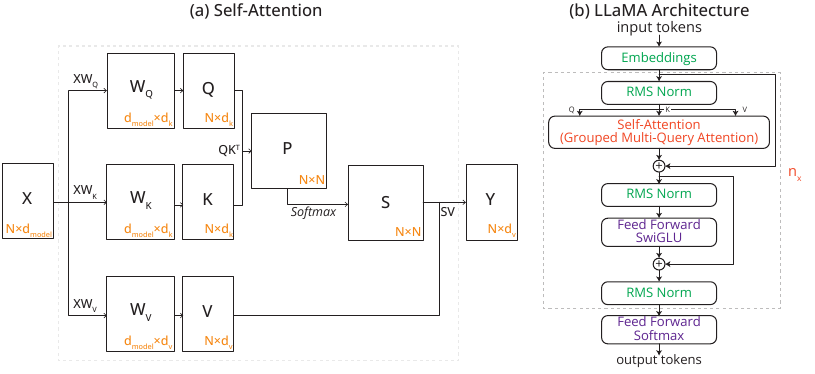}
	\caption{(a) Basic block diagram and calculations for the self-attention module. (b) LLaMA 2 model architecture diagram~\cite{touvron_llama_2023}.}
	\label{fig:attention}
\end{figure}

A self-attention module in Transformer-like architectures computes the attention weights and output values for a set of input vectors based on similarity.
The computation flow can be broken down into the following steps:
\begin{enumerate}[leftmargin=*]
	\item \textbf{Compute Query, Key, and Value vectors:} The input vectors (each token is a vector) are transformed into three sets of
	vectors: Query ($Q$), Key ($K$), and Value ($V$) vectors by multiplying weight matrices ($\mathbf{W_Q}$, $\mathbf{W_K}$, $\mathbf{W_V}$) to the input vectors using linear transformations.
	\item \textbf{Compute Attention Weights:} The attention weights are computed by taking the dot product of the
	Query vectors and Key vectors, then apply a softmax function to obtain probabilities. These
	probabilities (also called ``score'') represent the attention that should be paid to each input vector when computing
	the output values.
	\item \textbf{Compute Output Values:} The output values are computed by taking the weighted sum of the Value vectors, where the weights are determined by the attention probabilities calculated in the previous step.
	This is done for each input vector, resulting in a set of output vectors that have considered the
	relationships between all of the input vectors.
	\item \textbf{Apply Final Linear Transformation:} The output values are then passed through a final linear transformation (feed forward layer) to produce the final output vectors of the self-attention module.
\end{enumerate}

This computation flow allows the self-attention module to focus on different parts of the selectively input sequence, and compute output values that take into account the relationships between all of the
input vectors. 

Formally, given a set of query vectors $Q$, key vectors $K$, and value vectors $V$, the
Scale Dot-Product Attention can be defined as:
\begin{equation}
Attention(\mathbf{Q}, \mathbf{K}, \mathbf{V}) = softmax(\dfrac{\mathbf{Q}\mathbf{K}^\intercal}{\sqrt{d_k}})\mathbf{V}
\end{equation}

where $QK^T$ denotes the dot product between each query vector and each key vector, and
$\sqrt{d_k}$ is a scaling factor that helps to stabilize the gradient during training. The
resulting output has the same shape as the value vectors $V$, with the same number of
elements along each dimension.

\section{{\cn} Design}

\begin{figure}[h]
	\centering
	\includegraphics[width=0.7\columnwidth]{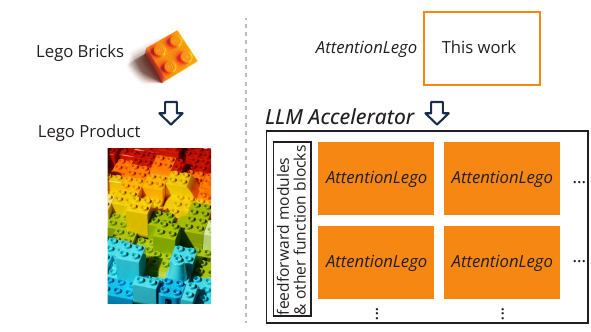}
	\caption{Core idea and the spatial scalability of  {\cn}.}
	\label{fig:idea}
\end{figure}

Fig.~\ref{fig:idea} illustrates the core idea of this work. This work implements a vanilla self-attention computation building block with Verilog HDL and the PIM macro behavioral model~\cite{yan_pislib_2024}. It can be conveniently tiled up for LLM with repetitive self-attention modules.

\subsection{Architecture}
\begin{figure}[h]
	\centering
	\includegraphics[width=1\columnwidth]{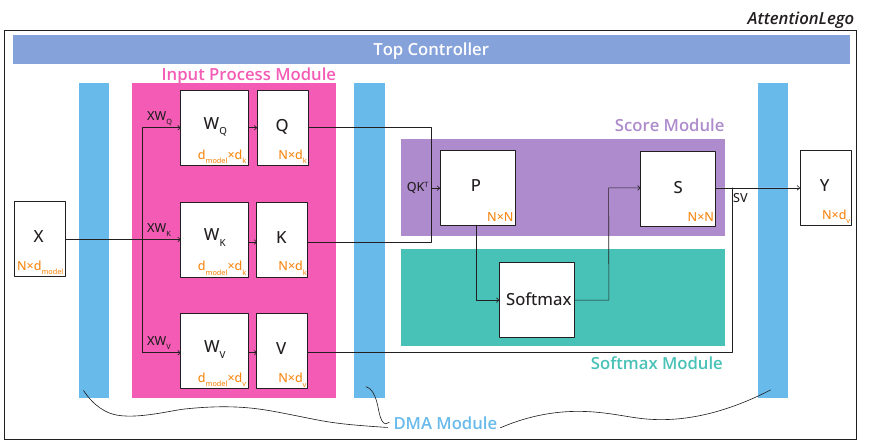}
	\caption{Architecture of {\cn}}.
	\label{fig:arc}
\end{figure}

As illustrated in Fig.~\ref{fig:arc}, {\cn} is divided into 5 parts:

\begin{table}[h]
\caption{{\cn} Modules}
\label{tab:arc}
\centering
\begin{tabular}{lll}
\toprule
No. & Module     & Description  \\
\midrule
1&Input Process module & compute $\mathbf{XW_Q}$, $\mathbf{XW_K}$, and $\mathbf{XW_V}$ \\
2&Score module & compute $\mathbf{QK}^\intercal$\\
3&Softmax module &  compute softmax nonlinear activation function in a vector manner\\
\multirow{2}{*}{4}&\multirow{2}{*}{DMA module} & controls data transfer among modules and between \\
&& {\cn} and external extra storage\\
5&top controller & controls the pipeline of computing of the above modules  \\
\bottomrule
\end{tabular}
\end{table}

\subsection{Input Process Module}
The Input Process module is responsible for (a) completing the writing and reading of the input weight matrices $\mathbf{W_Q}$, $\mathbf{W_K}$, and $\mathbf{W_V}$, and (b) execute the multiplying the input token with each of the three matrices.
The input process module works in 4 possible states: 
\begin{enumerate}[leftmargin=*]
	\item[a)]  IDLE: stand-by mode, do nothing;
	\item[b)]  WRITE: load pretrained the parameters of $\mathbf{W_Q}$, $\mathbf{W_K}$, and $\mathbf{W_V}$ into the input process module;
	\item[c)]  READ: read out the parameters of $\mathbf{W_Q}$, $\mathbf{W_K}$, and $\mathbf{W_V}$ for the testing and checking purpose;
	\item[d)]  CIM\footnote{We use ``CIM'' and ``PIM'' interchangeably.}: compute $\mathbf{Q}, \mathbf{K}, \mathbf{V}$ with the input $\mathbf{X}$ and the pre-loaded $\mathbf{W_Q}$, $\mathbf{W_K}$, and $\mathbf{W_V}$.
\end{enumerate}

The inputs ports of the input process module are: 
\begin{enumerate}[leftmargin=*]
	\item Data:
		\begin{enumerate}
			\item[a)] Weight data written/Input for CIM calculation: [DATA\_WIDTH * D\_MODEL-1:0] $data\_in$;
			\item[b)] Write to column address, range from 0 to 127, only for write mode: [ADDR\_COL\_WIDTH-1:0] $col\_sel$;
		\end{enumerate}
	\item Control:
		\begin{enumerate}
			\item[a)] Clock and reset signal, rising edge triggered: $clk$, $rst$;
			\item[b)] Chip selection signal, input is 1 for valid: $cs$;
			\item[c)] The mode selection is jointly determined by ($web$, $cimeb$): READ: (1,1), WRITE: (0,1), IDLE: (0,0), CIM: (1,0);
			\item[d)] Weight matrix selection signal: [2:0] $weight\_sel$. Among them, $\mathbf{Q}$: $weight\_sel=$(0,0); $\mathbf{K}$: $weight\_sel=$(0,1);$\mathbf{V}$: $weight\_sel=$(1,0);
		\end{enumerate}
\end{enumerate}

The output ports are:
\begin{enumerate}
	\item Data:
		\begin{enumerate}
			\item[a)] Output data for CIM mode: [DATA\_WIDTH * D\_k-1:0] $data\_out$;
			\item[b)] Read out the output data of (READ) mode: [DATA\_WIDTH * ` D\_MODEL-1:0] $mem\_data\_out$;
		\end{enumerate}
	\item Control:
	\begin{enumerate}
		\item[a)] Feedback completion calculation/read/write: $done$;
	\end{enumerate}
\end{enumerate}

The input process module consists of two parts, which are the APIM module for parameter storage and calculation, and the control module at the top for data distribution and state control.
The operating process is:
\begin{enumerate}
\item Writing weight matrix: Switch to WRITE mode and select a column of a specific matrix for writing each time. For the $\mathbf{Q}$ matrix, writing one column simultaneously requires 128 repetitions to complete, depending on the designed APIM IO bandwidth.
\item Multiplication of Input and Weight Matrix: Switch to CIM mode and feed the inputs (token) $\mathbf{X}$ parts by parts in order to match the stored weights/parameters. After passing in the data, the matrix-vector multiplication of $\mathbf{X}$ with each column vector in the weight matrix is carried out. It outputs a row vector with its length as $d_k$.
\item Reading out the weight matrix: used to test whether all weights are correctly written. Given the column selection signal, $col\_sel$ can read all rows of data for the 32 APIM modules in this column.
\end{enumerate}

Detailed of each working mode is given as follows:

WRITE mode:1
\begin{enumerate}
\item Decompose the input data into an array and write it in parallel to 32 APIM modules, with the writing position specified by $col\_sel$ decision. All are in IDLE state by default.
\item  For an APIM module, if the control signal is WRITE, it enters the BUSY state. In BUSY mode, repeat the following operation 128 times: select a row to write data at the given column address. Write the next row of data for this column in the next loop until all writes are completed.
\item After completing the write, switch to the DONE state to reset the state and transmit the DONE signal.
\end{enumerate}

READ mode:
\begin{enumerate}
\item According to $col\_sel$ and read weight data stored in 32 APIM modules in parallel.
\item For an APIM module, the control signal enters the BUSY state if the control signal is READ. In BUSY mode, repeat the following operation 128 times: select a row to read data at the given column address and store it in a temporary register. Read the next row of data for that column in the next loop until all readings are completed.
\item After completing the write, switch to the DONE state to reset the state and transmit the DONE signal.
\end{enumerate}

CIM mode:
\begin{enumerate}
\item The function is to organize and add the output results of a single APIM cycle, corresponding to the external circuit structure of the adder.
\item Due to the parallelism of the APIM module itself, data can be input, calculated, and read out in parallel. For a given column of addresses, it is necessary to repeat all calculations for that column 8 times. Due to the parallelism of the column output being 16, the above operation only needs to be repeated 8 times to complete the calculation for all columns. Therefore, completing a matrix multiplication requires 64 clock cycles.
\end{enumerate}

Description of the underlying APIM calculation module:
$\mathbf{Q}$, $\mathbf{K}$, and $\mathbf{V}$ each consisting of 32 APIM modules to store the entire parameters. Each APIM is a 128$\times$128 square matrix used to store weights and perform calculations. The stored weights are all 8-bit data. The input parallelism is 16, and a single input port shares 8 rows of addresses.The output parallelism is 16, a single output port shares 8 columns of addresses, and the ADC accuracy is 6 bits. Detailed can be found in \texttt{AttentionLego/InputProcess/src/defines.v}.

 \begin{figure}[h]
	\centering
	\includegraphics[width=0.4\columnwidth]{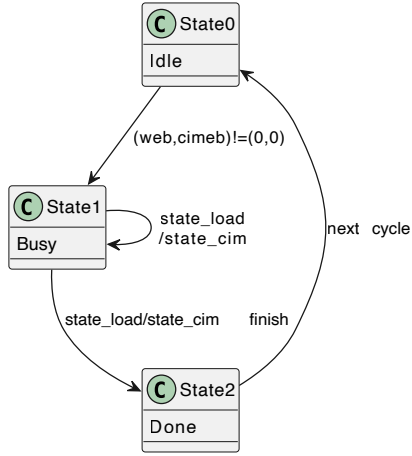}
	\caption{State-transfer diagram for the Input Process module.}
\end{figure}

\subsection{Score Module}

The Score module computes the score before Softmax shown in Fig.~\ref{fig:attention}, generating the square matrix $\mathbf{QK}^\intercal$. {\cn} design is to provide a template or starting point for fully customized Transformer accelerators. As exemplary dimensions, we choose the input is a vector with 128 elements that represents the value of a row in the $\mathbf{Q}$ or $\mathbf{K}^\intercal$ matrix, and the main output is a vector with 2048 elements that represents a row of the calculated $\mathbf{QK}^\intercal$. The detailed input and output ports are described as follows:
\begin{enumerate}[leftmargin=*]
\item Outputs:
\begin{itemize}
	\item $input\_done$: 1 bit, to indicate whether a row of $\mathbf{K}^\intercal$ as input vector has been cached to the inputs of this module;
	\item $output\_done$: 1 bit, to indicate whether calculation and transmission of a certain row of $\mathbf{QK}^\intercal$ has been done;
	\item $QK\_output$: 2048$\times$8 bits, to output a certain row of $\mathbf{QK}^\intercal$ (calculation results).
\end{itemize}
\item Inputs:
\begin{itemize}
	\item $clk$: clock;
	\item $cs$: chip select as the global enable for the Score module;
	\item $reset$: reset signal for the Score module;
	\item $K\_mode\_enable$: high active, controlling to start loading  $\mathbf{K}^\intercal$ row by row to the internal registers;
	\item $Q\_mode\_enable$:high active, controlling to start calculation with APIM for $\mathbf{QK}^\intercal$ row by row;
	\item $K\_address$: 11 bits, identify which row the input $\mathbf{K}^\intercal$ vector is in the $\mathbf{K}^\intercal$ matrix;
	\item $K\_input$: 128$\times$8 bits,  row vector inputs for the $\mathbf{K}^\intercal$;
	\item $Q\_input$: 128$\times$8 bits,  row vector inputs for the $\mathbf{Q}$;.
\end{itemize}
\end{enumerate}

\begin{figure}[h]
	\centering
	\includegraphics[width=0.6\columnwidth]{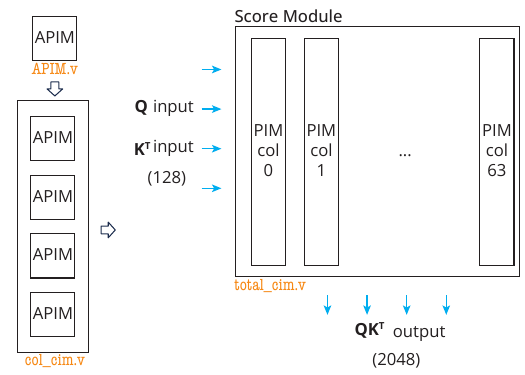}
	\caption{Architecture of the Score module}.
\end{figure}

This module comprises APIM modules of 32$\times$32 dimension (each APIM can store 32$\times$32 matrix of weights and perform general matrix-vector multiplication with 32$\times$1 input vector). 4 APIM modules are vertically arranged to form a 128$\times$32  matrix-vector multiplication engine, called \texttt{col\_cim} . Then we coalesce 64 columns of \texttt{col\_cim} to form a 128$\times$2048 PIM module to calculate 2048$\times$2048 $\mathbf{QK}^\intercal$.

This module has three states: State0=idle, State1=$K\_mode$, State2=$Q\_mode$. State1 and State2 are enabled by $K\_ mode\_enable$ and $Q\_mode\_enable$ signals. After each operation is completed in State1 or State2, the Score module returns to State0 and emits $input_done$ or $output_done$ signals to indicate the operations are done.

 \begin{figure}[h]
	\centering
	\includegraphics[width=1\columnwidth]{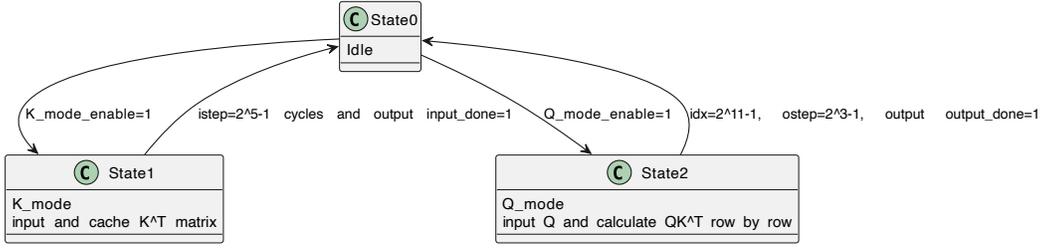}
	\caption{State-transfer diagram for the Score module.}
\end{figure}

 \begin{figure}[h]
	\centering
	\includegraphics[width=0.7\columnwidth]{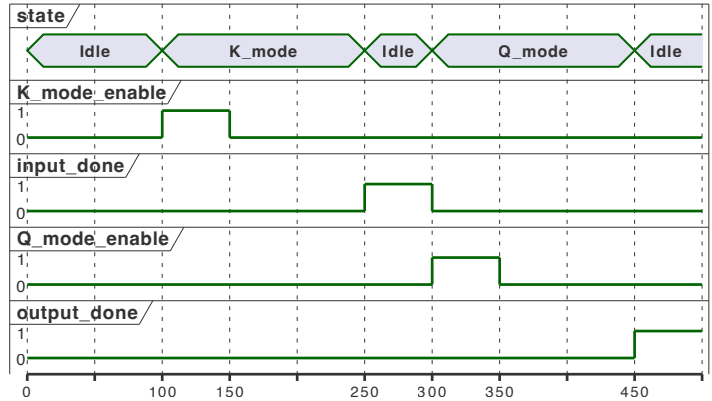}
	\caption{Timing diagram for the Score module.}
\end{figure}

\subsection{Softmax Module}
\textit{Softmax} module computes the softmax function, which is the core operation in the attention-based Transformer architecture.
The function of softmax is formulated as: given $\vec{v}=[v_1,v_2,\cdots,v_n]^\intercal$, output:
\begin{equation}
	softmax(\vec{v})=[a_1, a_2, \cdots, a_n]^\intercal, \mathrm{where\ } a_i=\frac{e^{v_i}}{\sum_{i=1}^{n}e^{v_i}  }
\end{equation}
This operation can be divided into two steps: first, find the exponents of each element, then do normalization.
The results of $\mathbf{QK}^\intercal$ block go through the Softmax block, which normalized all attention coefficients to 1.

This module, as explained in 2-steps, can be divided into two blocks:
\begin{enumerate}
	\item A look-up table implementation for the exponent function $exp\_function$: input is a 8-bit fixed-point number $x$, the output is a 16-bit fixed-point number $e^x$. We use a Python look-up table generator (\texttt{AttentionLego/Softmax/src/softmax.py}) to generate 256 possible cases for an 8-bit input $x$.
	\item Normalization block: here, we introduce a method to calculate the summation for all $e^x$ and normalization in two steps (clock cycles). The first cycle loads the inputs and computes the summation of all $e^x$; the second cycle calculates the normalization.
\end{enumerate}
In the example code, we provide a 32-number softmax implemented by digital logic.
 
 \begin{figure}[h]
 	\centering
 	\includegraphics[width=0.3\columnwidth]{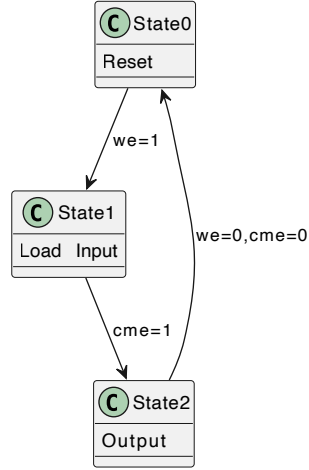}
 	\caption{State-transfer diagram for the Softmax module.}
 \end{figure}

\subsection{Direct Memory Access (DMA) Module}
{\cn} employs a special DMA module that orchestrates all of the data transportation, including the inputs and weights transfer between the {\cn} with external storage as well as the internal intermediate results.
DMA has three channels:
\begin{enumerate}
	\item The channel between external memory and the Input Process module. The DMA module converts the data read serially on the bus into parallel data and send it to the PIM module.
	\item The channel between the Input Process module and the Score module: feed the calculated $\mathbf{Q}$ or $\mathbf{K}^\intercal$ to the Score module.
	\item The channel between the Score module and the Softmax module.
\end{enumerate}

\begin{figure}[h]
	\centering
	\includegraphics[width=0.6\columnwidth]{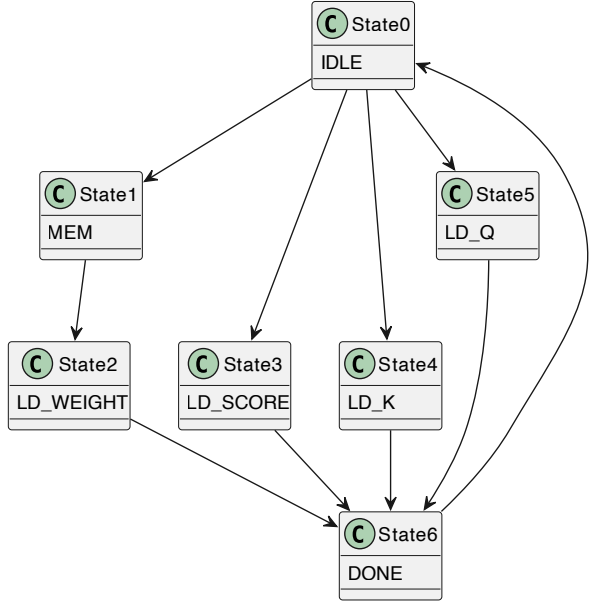}
	\caption{State-transfer diagram for the DMA module.}
	\label{fig:dma_statediagram}
\end{figure}

Fig.~\ref{fig:dma_statediagram} illustrates the state transfer diagram for the DMA module. Admittedly the data loading control for the Input Process module, the Score module and the Softmax module can be fused into those modules; here we make the DMA module a standalone one for the purpose of adapting the {\cn} scheme for various on-chip and off-chip data transfer bandwidth.

\subsection{Top Controller Module}
This module is mainly responsible for managing and coordinating communication, data flow, and functional operations between different modules within the chip to ensure that the entire system can operate normally according to design requirements. This design implements the inference operation of the transformer's attention module in the case of a batch size of 1. The entire process includes each module taking weights from memory and calculating the $\mathbf{Q},\mathbf{K},\mathbf{V}$ matrix, attention score, and softmax. The top controller controls a module to start working through the enable signal. After the module finishes working, it sends a do signal to the top controller to inform them that the work is complete so that the top controller can continue to control subsequent work. The top controller completed the design of the entire process timing logic through different enable and do signals.
The main output is the enable signal for controlling the selection and calculation of the input process module, the enable signal for controlling the selection, input, and calculation of the attention score calculation module, the enable signal for controlling the selection and calculation of the softmax calculation module, and the enable signal for controlling the DMA transmission information. The main input is the corresponding done signal.

This module consists of a two-layer nested state machine. The outer state machine has four states, with state 0 being the ready state and containing four inner states. It mainly performs weight loading, input loading, k-matrix calculation and loading, and the calculation of the q-vector of the first token. The last three states are the main part of inference, with each loop performing a token inference. State 1 transfers the q vector of the current token to the attention score calculation module, State 2 calculates the attention score, the q vector of the new token is calculated (not required for the last loop), and the softmax value of the previous token is calculated (not needed in the first loop), State 3 sends the calculated attention score of the current token to the softmax calculation module.
The inner state machine has four states that execute sequentially. State 0.0 loads weight from memory into the input process module, state 0.1 loads input from memory into the input process module, state 0.2 completes the calculation of the $\mathbf{K}$ matrix in the input process module, state 0.3 loads the $\mathbf{K}$ matrix into the attention score calculation module, and completes the calculation of the q-vector of the first token.

The Top controller consists of a two-layer nested state machine, with state 0 in the outer state being the ready state and states 1, 2, and 3 forming a loop. Each loop completes the calculation for a token, and state 0 contains an inner state machine with four states. The four states proceed sequentially to complete the preparation phase. The Top controller outputs an enable signal in each stage, and each time it receives a done signal, it enters the next stage.

The Top controller uses a pipeline to complete token-level parallel inference during the inference process. In the second state of each loop, the attention score calculation module processes the current token, the softmax calculation module calculates the previous token, the input process module calculates the q vector of the next token, and during DMA transmission, the three calculation modules temporarily stop computing, This can maximize hardware utilization while ensuring the accuracy of computation and transmission.

%
%
%
%
%

\section{Conclusion and Perspective}
This paper reveals the basic design of {\cn}. This work develops a vanilla self-attention module in Verilog HDL based on the PIM macro behavioral model. Thanks to the duality of efficient storage and computation of PIM macro for tensor processing, this work deploys the matrix multiplications in LLMs onto a PIM-based weight-stationary data flow. In this process, the parameters of LLMs are loaded into {\cn} only once. This is the major power and energy-saving technique used in this work. This work, for the moment, releases the initial version of the source code; more quantitative analysis with the proposed {\cn} method and design are coming up.

\begin{ack}
This work was supported by the National Key R\&D Program of China (2023YFB4502200), and National Natural Science Foundation of China (92264201, 92364102, T2350006), by the 111 Project under Grant B18001. Partial of this work is the results out of an undergraduate course at Peking University with course no. 04835370 in 2023'fall semester.
\end{ack}

\bibliographystyle{unsrtnat}
\bibliography{ref}

\begin{thebibliography}{12}
\providecommand{\natexlab}[1]{#1}
\providecommand{\url}[1]{\texttt{#1}}
\expandafter\ifx\csname urlstyle\endcsname\relax
  \providecommand{\doi}[1]{doi: #1}\else
  \providecommand{\doi}{doi: \begingroup \urlstyle{rm}\Url}\fi

\bibitem[Touvron et~al.(2023{\natexlab{a}})Touvron, Lavril, Izacard, Martinet,
  Lachaux, Lacroix, Rozière, Goyal, Hambro, Azhar, Rodriguez, Joulin, Grave,
  and Lample]{touvron_llama_2023-1}
Hugo Touvron, Thibaut Lavril, Gautier Izacard, Xavier Martinet, Marie-Anne
  Lachaux, Timothée Lacroix, Baptiste Rozière, Naman Goyal, Eric Hambro,
  Faisal Azhar, Aurelien Rodriguez, Armand Joulin, Edouard Grave, and Guillaume
  Lample.
\newblock {LLaMA}: {Open} and {Efficient} {Foundation} {Language} {Models},
  February 2023{\natexlab{a}}.
\newblock URL \url{http://arxiv.org/abs/2302.13971}.
\newblock arXiv:2302.13971 [cs].

\bibitem[Touvron et~al.(2023{\natexlab{b}})Touvron, Martin, Stone, Albert,
  Almahairi, Babaei, Bashlykov, Batra, Bhargava, Bhosale, Bikel, Blecher,
  Ferrer, Chen, Cucurull, Esiobu, Fernandes, Fu, Fu, Fuller, Gao, Goswami,
  Goyal, Hartshorn, Hosseini, Hou, Inan, Kardas, Kerkez, Khabsa, Kloumann,
  Korenev, Koura, Lachaux, Lavril, Lee, Liskovich, Lu, Mao, Martinet, Mihaylov,
  Mishra, Molybog, Nie, Poulton, Reizenstein, Rungta, Saladi, Schelten, Silva,
  Smith, Subramanian, Tan, Tang, Taylor, Williams, Kuan, Xu, Yan, Zarov, Zhang,
  Fan, Kambadur, Narang, Rodriguez, Stojnic, Edunov, and
  Scialom]{touvron_llama_2023}
Hugo Touvron, Louis Martin, Kevin Stone, Peter Albert, Amjad Almahairi, Yasmine
  Babaei, Nikolay Bashlykov, Soumya Batra, Prajjwal Bhargava, Shruti Bhosale,
  Dan Bikel, Lukas Blecher, Cristian~Canton Ferrer, Moya Chen, Guillem
  Cucurull, David Esiobu, Jude Fernandes, Jeremy Fu, Wenyin Fu, Brian Fuller,
  Cynthia Gao, Vedanuj Goswami, Naman Goyal, Anthony Hartshorn, Saghar
  Hosseini, Rui Hou, Hakan Inan, Marcin Kardas, Viktor Kerkez, Madian Khabsa,
  Isabel Kloumann, Artem Korenev, Punit~Singh Koura, Marie-Anne Lachaux,
  Thibaut Lavril, Jenya Lee, Diana Liskovich, Yinghai Lu, Yuning Mao, Xavier
  Martinet, Todor Mihaylov, Pushkar Mishra, Igor Molybog, Yixin Nie, Andrew
  Poulton, Jeremy Reizenstein, Rashi Rungta, Kalyan Saladi, Alan Schelten, Ruan
  Silva, Eric~Michael Smith, Ranjan Subramanian, Xiaoqing~Ellen Tan, Binh Tang,
  Ross Taylor, Adina Williams, Jian~Xiang Kuan, Puxin Xu, Zheng Yan, Iliyan
  Zarov, Yuchen Zhang, Angela Fan, Melanie Kambadur, Sharan Narang, Aurelien
  Rodriguez, Robert Stojnic, Sergey Edunov, and Thomas Scialom.
\newblock Llama 2: {Open} {Foundation} and {Fine}-{Tuned} {Chat} {Models}, July
  2023{\natexlab{b}}.
\newblock URL \url{http://arxiv.org/abs/2307.09288}.
\newblock arXiv:2307.09288 [cs].

\bibitem[Workshop et~al.(2023)Workshop, Scao, Fan, Akiki, Pavlick, Ilić,
  Hesslow, Castagné, Luccioni, Yvon, Gallé, Tow, Rush, Biderman, Webson,
  Ammanamanchi, Wang, Sagot, Muennighoff, del Moral, Ruwase, Bawden, Bekman,
  McMillan-Major, Beltagy, Nguyen, Saulnier, Tan, Suarez, Sanh, Laurençon,
  Jernite, Launay, Mitchell, Raffel, Gokaslan, Simhi, Soroa, Aji, Alfassy,
  Rogers, Nitzav, Xu, Mou, Emezue, Klamm, Leong, van Strien, Adelani, Radev,
  Ponferrada, Levkovizh, Kim, Natan, De~Toni, Dupont, Kruszewski, Pistilli,
  Elsahar, Benyamina, Tran, Yu, Abdulmumin, Johnson, Gonzalez-Dios, de~la Rosa,
  Chim, Dodge, Zhu, Chang, Frohberg, Tobing, Bhattacharjee, Almubarak, Chen,
  Lo, Von~Werra, Weber, Phan, allal, Tanguy, Dey, Muñoz, Masoud, Grandury,
  Šaško, Huang, Coavoux, Singh, Jiang, Vu, Jauhar, Ghaleb, Subramani,
  Kassner, Khamis, Nguyen, Espejel, de~Gibert, Villegas, Henderson, Colombo,
  Amuok, Lhoest, Harliman, Bommasani, López, Ribeiro, Osei, Pyysalo, Nagel,
  Bose, Muhammad, Sharma, Longpre, Nikpoor, Silberberg, Pai, Zink, Torrent,
  Schick, Thrush, Danchev, Nikoulina, Laippala, Lepercq, Prabhu, Alyafeai,
  Talat, Raja, Heinzerling, Si, Taşar, Salesky, Mielke, Lee, Sharma, Santilli,
  Chaffin, Stiegler, Datta, Szczechla, Chhablani, Wang, Pandey, Strobelt,
  Fries, Rozen, Gao, Sutawika, Bari, Al-shaibani, Manica, Nayak, Teehan,
  Albanie, Shen, Ben-David, Bach, Kim, Bers, Fevry, Neeraj, Thakker, Raunak,
  Tang, Yong, Sun, Brody, Uri, Tojarieh, Roberts, Chung, Tae, Phang, Press, Li,
  Narayanan, Bourfoune, Casper, Rasley, Ryabinin, Mishra, Zhang, Shoeybi,
  Peyrounette, Patry, Tazi, Sanseviero, von Platen, Cornette, Lavallée,
  Lacroix, Rajbhandari, Gandhi, Smith, Requena, Patil, Dettmers, Baruwa, Singh,
  Cheveleva, Ligozat, Subramonian, Névéol, Lovering, Garrette, Tunuguntla,
  Reiter, Taktasheva, Voloshina, Bogdanov, Winata, Schoelkopf, Kalo, Novikova,
  Forde, Clive, Kasai, Kawamura, Hazan, Carpuat, Clinciu, Kim, Cheng, Serikov,
  Antverg, van~der Wal, Zhang, Zhang, Gehrmann, Mirkin, Pais, Shavrina,
  Scialom, Yun, Limisiewicz, Rieser, Protasov, Mikhailov, Pruksachatkun,
  Belinkov, Bamberger, Kasner, Rueda, Pestana, Feizpour, Khan, Faranak, Santos,
  Hevia, Unldreaj, Aghagol, Abdollahi, Tammour, HajiHosseini, Behroozi,
  Ajibade, Saxena, Ferrandis, McDuff, Contractor, Lansky, David, Kiela, Nguyen,
  Tan, Baylor, Ozoani, Mirza, Ononiwu, Rezanejad, Jones, Bhattacharya,
  Solaiman, Sedenko, Nejadgholi, Passmore, Seltzer, Sanz, Dutra, Samagaio,
  Elbadri, Mieskes, Gerchick, Akinlolu, McKenna, Qiu, Ghauri, Burynok, Abrar,
  Rajani, Elkott, Fahmy, Samuel, An, Kromann, Hao, Alizadeh, Shubber, Wang,
  Roy, Viguier, Le, Oyebade, Le, Yang, Nguyen, Kashyap, Palasciano, Callahan,
  Shukla, Miranda-Escalada, Singh, Beilharz, Wang, Brito, Zhou, Jain, Xu,
  Fourrier, Periñán, Molano, Yu, Manjavacas, Barth, Fuhrimann, Altay, Bayrak,
  Burns, Vrabec, Bello, Dash, Kang, Giorgi, Golde, Posada, Sivaraman,
  Bulchandani, Liu, Shinzato, de~Bykhovetz, Takeuchi, Pàmies, Castillo,
  Nezhurina, Sänger, Samwald, Cullan, Weinberg, De~Wolf, Mihaljcic, Liu,
  Freidank, Kang, Seelam, Dahlberg, Broad, Muellner, Fung, Haller,
  Chandrasekhar, Eisenberg, Martin, Canalli, Su, Su, Cahyawijaya, Garda,
  Deshmukh, Mishra, Kiblawi, Ott, Sang-aroonsiri, Kumar, Schweter, Bharati,
  Laud, Gigant, Kainuma, Kusa, Labrak, Bajaj, Venkatraman, Xu, Xu, Xu, Tan,
  Xie, Ye, Bras, Belkada, and Wolf]{workshop_bloom_2023}
BigScience Workshop, Teven~Le Scao, Angela Fan, Christopher Akiki, Ellie
  Pavlick, Suzana Ilić, Daniel Hesslow, Roman Castagné, Alexandra~Sasha
  Luccioni, François Yvon, Matthias Gallé, Jonathan Tow, Alexander~M. Rush,
  Stella Biderman, Albert Webson, Pawan~Sasanka Ammanamanchi, Thomas Wang,
  Benoît Sagot, Niklas Muennighoff, Albert~Villanova del Moral, Olatunji
  Ruwase, Rachel Bawden, Stas Bekman, Angelina McMillan-Major, Iz~Beltagy, Huu
  Nguyen, Lucile Saulnier, Samson Tan, Pedro~Ortiz Suarez, Victor Sanh, Hugo
  Laurençon, Yacine Jernite, Julien Launay, Margaret Mitchell, Colin Raffel,
  Aaron Gokaslan, Adi Simhi, Aitor Soroa, Alham~Fikri Aji, Amit Alfassy, Anna
  Rogers, Ariel~Kreisberg Nitzav, Canwen Xu, Chenghao Mou, Chris Emezue,
  Christopher Klamm, Colin Leong, Daniel van Strien, David~Ifeoluwa Adelani,
  Dragomir Radev, Eduardo~González Ponferrada, Efrat Levkovizh, Ethan Kim,
  Eyal~Bar Natan, Francesco De~Toni, Gérard Dupont, Germán Kruszewski, Giada
  Pistilli, Hady Elsahar, Hamza Benyamina, Hieu Tran, Ian Yu, Idris Abdulmumin,
  Isaac Johnson, Itziar Gonzalez-Dios, Javier de~la Rosa, Jenny Chim, Jesse
  Dodge, Jian Zhu, Jonathan Chang, Jörg Frohberg, Joseph Tobing, Joydeep
  Bhattacharjee, Khalid Almubarak, Kimbo Chen, Kyle Lo, Leandro Von~Werra, Leon
  Weber, Long Phan, Loubna~Ben allal, Ludovic Tanguy, Manan Dey, Manuel~Romero
  Muñoz, Maraim Masoud, María Grandury, Mario Šaško, Max Huang, Maximin
  Coavoux, Mayank Singh, Mike Tian-Jian Jiang, Minh~Chien Vu, Mohammad~A.
  Jauhar, Mustafa Ghaleb, Nishant Subramani, Nora Kassner, Nurulaqilla Khamis,
  Olivier Nguyen, Omar Espejel, Ona de~Gibert, Paulo Villegas, Peter Henderson,
  Pierre Colombo, Priscilla Amuok, Quentin Lhoest, Rheza Harliman, Rishi
  Bommasani, Roberto~Luis López, Rui Ribeiro, Salomey Osei, Sampo Pyysalo,
  Sebastian Nagel, Shamik Bose, Shamsuddeen~Hassan Muhammad, Shanya Sharma,
  Shayne Longpre, Somaieh Nikpoor, Stanislav Silberberg, Suhas Pai, Sydney
  Zink, Tiago~Timponi Torrent, Timo Schick, Tristan Thrush, Valentin Danchev,
  Vassilina Nikoulina, Veronika Laippala, Violette Lepercq, Vrinda Prabhu, Zaid
  Alyafeai, Zeerak Talat, Arun Raja, Benjamin Heinzerling, Chenglei Si,
  Davut~Emre Taşar, Elizabeth Salesky, Sabrina~J. Mielke, Wilson~Y. Lee,
  Abheesht Sharma, Andrea Santilli, Antoine Chaffin, Arnaud Stiegler, Debajyoti
  Datta, Eliza Szczechla, Gunjan Chhablani, Han Wang, Harshit Pandey, Hendrik
  Strobelt, Jason~Alan Fries, Jos Rozen, Leo Gao, Lintang Sutawika, M.~Saiful
  Bari, Maged~S. Al-shaibani, Matteo Manica, Nihal Nayak, Ryan Teehan, Samuel
  Albanie, Sheng Shen, Srulik Ben-David, Stephen~H. Bach, Taewoon Kim, Tali
  Bers, Thibault Fevry, Trishala Neeraj, Urmish Thakker, Vikas Raunak, Xiangru
  Tang, Zheng-Xin Yong, Zhiqing Sun, Shaked Brody, Yallow Uri, Hadar Tojarieh,
  Adam Roberts, Hyung~Won Chung, Jaesung Tae, Jason Phang, Ofir Press, Conglong
  Li, Deepak Narayanan, Hatim Bourfoune, Jared Casper, Jeff Rasley, Max
  Ryabinin, Mayank Mishra, Minjia Zhang, Mohammad Shoeybi, Myriam Peyrounette,
  Nicolas Patry, Nouamane Tazi, Omar Sanseviero, Patrick von Platen, Pierre
  Cornette, Pierre~François Lavallée, Rémi Lacroix, Samyam Rajbhandari,
  Sanchit Gandhi, Shaden Smith, Stéphane Requena, Suraj Patil, Tim Dettmers,
  Ahmed Baruwa, Amanpreet Singh, Anastasia Cheveleva, Anne-Laure Ligozat, Arjun
  Subramonian, Aurélie Névéol, Charles Lovering, Dan Garrette, Deepak
  Tunuguntla, Ehud Reiter, Ekaterina Taktasheva, Ekaterina Voloshina, Eli
  Bogdanov, Genta~Indra Winata, Hailey Schoelkopf, Jan-Christoph Kalo,
  Jekaterina Novikova, Jessica~Zosa Forde, Jordan Clive, Jungo Kasai, Ken
  Kawamura, Liam Hazan, Marine Carpuat, Miruna Clinciu, Najoung Kim, Newton
  Cheng, Oleg Serikov, Omer Antverg, Oskar van~der Wal, Rui Zhang, Ruochen
  Zhang, Sebastian Gehrmann, Shachar Mirkin, Shani Pais, Tatiana Shavrina,
  Thomas Scialom, Tian Yun, Tomasz Limisiewicz, Verena Rieser, Vitaly Protasov,
  Vladislav Mikhailov, Yada Pruksachatkun, Yonatan Belinkov, Zachary Bamberger,
  Zdeněk Kasner, Alice Rueda, Amanda Pestana, Amir Feizpour, Ammar Khan, Amy
  Faranak, Ana Santos, Anthony Hevia, Antigona Unldreaj, Arash Aghagol, Arezoo
  Abdollahi, Aycha Tammour, Azadeh HajiHosseini, Bahareh Behroozi, Benjamin
  Ajibade, Bharat Saxena, Carlos~Muñoz Ferrandis, Daniel McDuff, Danish
  Contractor, David Lansky, Davis David, Douwe Kiela, Duong~A. Nguyen, Edward
  Tan, Emi Baylor, Ezinwanne Ozoani, Fatima Mirza, Frankline Ononiwu, Habib
  Rezanejad, Hessie Jones, Indrani Bhattacharya, Irene Solaiman, Irina Sedenko,
  Isar Nejadgholi, Jesse Passmore, Josh Seltzer, Julio~Bonis Sanz, Livia Dutra,
  Mairon Samagaio, Maraim Elbadri, Margot Mieskes, Marissa Gerchick, Martha
  Akinlolu, Michael McKenna, Mike Qiu, Muhammed Ghauri, Mykola Burynok, Nafis
  Abrar, Nazneen Rajani, Nour Elkott, Nour Fahmy, Olanrewaju Samuel, Ran An,
  Rasmus Kromann, Ryan Hao, Samira Alizadeh, Sarmad Shubber, Silas Wang, Sourav
  Roy, Sylvain Viguier, Thanh Le, Tobi Oyebade, Trieu Le, Yoyo Yang, Zach
  Nguyen, Abhinav~Ramesh Kashyap, Alfredo Palasciano, Alison Callahan, Anima
  Shukla, Antonio Miranda-Escalada, Ayush Singh, Benjamin Beilharz, Bo~Wang,
  Caio Brito, Chenxi Zhou, Chirag Jain, Chuxin Xu, Clémentine Fourrier,
  Daniel~León Periñán, Daniel Molano, Dian Yu, Enrique Manjavacas, Fabio
  Barth, Florian Fuhrimann, Gabriel Altay, Giyaseddin Bayrak, Gully Burns,
  Helena~U. Vrabec, Imane Bello, Ishani Dash, Jihyun Kang, John Giorgi, Jonas
  Golde, Jose~David Posada, Karthik~Rangasai Sivaraman, Lokesh Bulchandani,
  Lu~Liu, Luisa Shinzato, Madeleine~Hahn de~Bykhovetz, Maiko Takeuchi, Marc
  Pàmies, Maria~A. Castillo, Marianna Nezhurina, Mario Sänger, Matthias
  Samwald, Michael Cullan, Michael Weinberg, Michiel De~Wolf, Mina Mihaljcic,
  Minna Liu, Moritz Freidank, Myungsun Kang, Natasha Seelam, Nathan Dahlberg,
  Nicholas~Michio Broad, Nikolaus Muellner, Pascale Fung, Patrick Haller, Ramya
  Chandrasekhar, Renata Eisenberg, Robert Martin, Rodrigo Canalli, Rosaline Su,
  Ruisi Su, Samuel Cahyawijaya, Samuele Garda, Shlok~S. Deshmukh, Shubhanshu
  Mishra, Sid Kiblawi, Simon Ott, Sinee Sang-aroonsiri, Srishti Kumar, Stefan
  Schweter, Sushil Bharati, Tanmay Laud, Théo Gigant, Tomoya Kainuma, Wojciech
  Kusa, Yanis Labrak, Yash~Shailesh Bajaj, Yash Venkatraman, Yifan Xu, Yingxin
  Xu, Yu~Xu, Zhe Tan, Zhongli Xie, Zifan Ye, Mathilde Bras, Younes Belkada, and
  Thomas Wolf.
\newblock {BLOOM}: {A} {176B}-{Parameter} {Open}-{Access} {Multilingual}
  {Language} {Model}, June 2023.
\newblock URL \url{http://arxiv.org/abs/2211.05100}.
\newblock arXiv:2211.05100 [cs].

\bibitem[Dey et~al.(2023)Dey, Gosal, Zhiming, Chen, Khachane, Marshall,
  Pathria, Tom, and Hestness]{dey_cerebras-gpt_2023}
Nolan Dey, Gurpreet Gosal, Zhiming, Chen, Hemant Khachane, William Marshall,
  Ribhu Pathria, Marvin Tom, and Joel Hestness.
\newblock Cerebras-{GPT}: {Open} {Compute}-{Optimal} {Language} {Models}
  {Trained} on the {Cerebras} {Wafer}-{Scale} {Cluster}, April 2023.
\newblock URL \url{http://arxiv.org/abs/2304.03208}.
\newblock arXiv:2304.03208 [cs].

\bibitem[Dao et~al.(2022)Dao, Fu, Ermon, Rudra, and
  Ré]{dao_flashattention_2022}
Tri Dao, Daniel~Y. Fu, Stefano Ermon, Atri Rudra, and Christopher Ré.
\newblock {FlashAttention}: {Fast} and {Memory}-{Efficient} {Exact} {Attention}
  with {IO}-{Awareness}, June 2022.
\newblock URL \url{http://arxiv.org/abs/2205.14135}.
\newblock arXiv:2205.14135 [cs].

\bibitem[Black et~al.(2022)Black, Biderman, Hallahan, Anthony, Gao, Golding,
  He, Leahy, McDonell, Phang, Pieler, Prashanth, Purohit, Reynolds, Tow, Wang,
  and Weinbach]{black_gpt-neox-20b_2022}
Sid Black, Stella Biderman, Eric Hallahan, Quentin Anthony, Leo Gao, Laurence
  Golding, Horace He, Connor Leahy, Kyle McDonell, Jason Phang, Michael Pieler,
  USVSN~Sai Prashanth, Shivanshu Purohit, Laria Reynolds, Jonathan Tow, Ben
  Wang, and Samuel Weinbach.
\newblock {GPT}-{NeoX}-{20B}: {An} {Open}-{Source} {Autoregressive} {Language}
  {Model}, April 2022.
\newblock URL \url{http://arxiv.org/abs/2204.06745}.
\newblock arXiv:2204.06745 [cs].

\bibitem[Li et~al.(2023)Li, Bubeck, Eldan, Del~Giorno, Gunasekar, and
  Lee]{li_textbooks_2023}
Yuanzhi Li, Sébastien Bubeck, Ronen Eldan, Allie Del~Giorno, Suriya Gunasekar,
  and Yin~Tat Lee.
\newblock Textbooks {Are} {All} {You} {Need} {II}: phi-1.5 technical report,
  September 2023.
\newblock URL \url{http://arxiv.org/abs/2309.05463}.
\newblock arXiv:2309.05463 [cs].

\bibitem[Biderman et~al.(2023)Biderman, Schoelkopf, Anthony, Bradley, O'Brien,
  Hallahan, Khan, Purohit, Prashanth, Raff, Skowron, Sutawika, and van~der
  Wal]{biderman_pythia_2023}
Stella Biderman, Hailey Schoelkopf, Quentin Anthony, Herbie Bradley, Kyle
  O'Brien, Eric Hallahan, Mohammad~Aflah Khan, Shivanshu Purohit, USVSN~Sai
  Prashanth, Edward Raff, Aviya Skowron, Lintang Sutawika, and Oskar van~der
  Wal.
\newblock Pythia: {A} {Suite} for {Analyzing} {Large} {Language} {Models}
  {Across} {Training} and {Scaling}, May 2023.
\newblock URL \url{http://arxiv.org/abs/2304.01373}.
\newblock arXiv:2304.01373 [cs].

\bibitem[Vaswani et~al.(2023)Vaswani, Shazeer, Parmar, Uszkoreit, Jones, Gomez,
  Kaiser, and Polosukhin]{vaswani_attention_2023}
Ashish Vaswani, Noam Shazeer, Niki Parmar, Jakob Uszkoreit, Llion Jones,
  Aidan~N. Gomez, Lukasz Kaiser, and Illia Polosukhin.
\newblock Attention {Is} {All} {You} {Need}, August 2023.
\newblock URL \url{http://arxiv.org/abs/1706.03762}.
\newblock arXiv:1706.03762 [cs].

\bibitem[Yan et~al.(2019{\natexlab{a}})Yan, Li, Qiao, Xue, Chang, Chen, and
  Li]{yan_resistive_2019}
Bonan Yan, Bing Li, Ximing Qiao, Cheng-Xin Xue, Meng‐Fan Chang, Yiran Chen,
  and Hai~(Helen) Li.
\newblock Resistive {Memory}‐{Based} {In}‐{Memory} {Computing}: {From}
  {Device} and {Large}‐{Scale} {Integration} {System} {Perspectives}.
\newblock \emph{Advanced Intelligent Systems}, 1\penalty0 (7):\penalty0
  1900068, November 2019{\natexlab{a}}.
\newblock ISSN 2640-4567, 2640-4567.
\newblock \doi{10.1002/aisy.201900068}.
\newblock URL \url{https://onlinelibrary.wiley.com/doi/10.1002/aisy.201900068}.

\bibitem[Yan et~al.(2019{\natexlab{b}})Yan, Yang, Chen, Chang, Su, Hsu, Li,
  Lee, Sheu, Ho, Wu, Chang, Chen, and Li]{yan_rram-based_2019}
Bonan Yan, Qing Yang, Wei-Hao Chen, Kung-Tang Chang, Jian-Wei Su, Chien-Hua
  Hsu, Sih-Han Li, Heng-Yuan Lee, Shyh-Shyuan Sheu, Mon-Shu Ho, Qing Wu,
  Meng-Fan Chang, Yiran Chen, and Hai Li.
\newblock {RRAM}-based {Spiking} {Nonvolatile} {Computing}-{In}-{Memory}
  {Processing} {Engine} with {Precision}-{Configurable} {In} {Situ} {Nonlinear}
  {Activation}.
\newblock In \emph{2019 {Symposium} on {VLSI} {Technology}}, pages T86--T87,
  Kyoto, Japan, June 2019{\natexlab{b}}. IEEE.
\newblock ISBN 978-4-86348-719-2.
\newblock \doi{10.23919/VLSIT.2019.8776485}.
\newblock URL \url{https://ieeexplore.ieee.org/document/8776485/}.

\bibitem[Yan(2024)]{yan_pislib_2024}
Bonan Yan.
\newblock {PISLIB}, January 2024.
\newblock URL \url{https://bonany.gitlab.io/pis/}.

\end{thebibliography}

\end{document}